
\font\twelverm=cmr10  scaled 1200   \font\twelvei=cmmi10  scaled 1200
\font\twelvesy=cmsy10 scaled 1200   \font\twelveex=cmex10 scaled 1200
\font\twelvebf=cmbx10 scaled 1200   \font\twelvesl=cmsl10 scaled 1200
\font\twelvett=cmtt10 scaled 1200   \font\twelveit=cmti10 scaled 1200
\font\twelvesc=cmcsc10 scaled 1200  \font\twelvesf=cmss10 scaled 1200
\skewchar\twelvei='177   \skewchar\twelvesy='60


\def\twelvepoint{\normalbaselineskip=12.4pt plus 0.1pt minus 0.1pt
  \abovedisplayskip 12.4pt plus 3pt minus 9pt
  \belowdisplayskip 12.4pt plus 3pt minus 9pt
  \abovedisplayshortskip 0pt plus 3pt
  \belowdisplayshortskip 7.2pt plus 3pt minus 4pt
  \smallskipamount=3.6pt plus1.2pt minus1.2pt
  \medskipamount=7.2pt plus2.4pt minus2.4pt
  \bigskipamount=14.4pt plus4.8pt minus4.8pt
  \def\rm{\fam0\twelverm}          \def\it{\fam\itfam\twelveit}%
  \def\sl{\fam\slfam\twelvesl}     \def\bf{\fam\bffam\twelvebf}%
  \def\mit{\fam 1}                 \def\cal{\fam 2}%
  \def\sc{\twelvesc}               \def\tt{\twelvett}
  \def\sf{\twelvesf}
  \textfont0=\twelverm   \scriptfont0=\tenrm   \scriptscriptfont0=\sevenrm
  \textfont1=\twelvei    \scriptfont1=\teni    \scriptscriptfont1=\seveni
  \textfont2=\twelvesy   \scriptfont2=\tensy   \scriptscriptfont2=\sevensy
  \textfont3=\twelveex   \scriptfont3=\twelveex  \scriptscriptfont3=\twelveex
  \textfont\itfam=\twelveit
  \textfont\slfam=\twelvesl
  \textfont\bffam=\twelvebf \scriptfont\bffam=\tenbf
  \scriptscriptfont\bffam=\sevenbf
  \normalbaselines\rm}



\def\beginlinemode{\endmode
  \begingroup\parskip=0pt \obeylines\def\\{\par}\def\endmode{\par\endgroup}}
\def\beginparmode{\endmode
  \begingroup \def\endmode{\par\endgroup}}
\let\endmode=\par
{\obeylines\gdef\
{}}
\def\singlespace{\baselineskip=\normalbaselineskip}

\def\oneandahalfspace{\baselineskip=\normalbaselineskip
  \multiply\baselineskip by 3 \divide\baselineskip by 2}
\def\doublespace{\baselineskip=\normalbaselineskip \multiply\baselineskip by 2}

\newcount\firstpageno
\firstpageno=2
\footline={\ifnum\pageno<\firstpageno{\hfil}\else{\hfil\twelverm\folio\hfil}\fi}
\def\toppageno{\global\footline={\hfil}\global\headline
  ={\ifnum\pageno<\firstpageno{\hfil}\else{\hfil\twelverm\folio\hfil}\fi}}
\let\rawfootnote=\footnote              
\def\footnote#1#2{{\rm\singlespace\parindent=0pt\parskip=0pt
  \rawfootnote{#1}{#2\hfill\vrule height 0pt depth 6pt width 0pt}}}
\def\raggedcenter{\leftskip=4em plus 12em \rightskip=\leftskip
  \parindent=0pt \parfillskip=0pt \spaceskip=.3333em \xspaceskip=.5em
  \pretolerance=9999 \tolerance=9999
  \hyphenpenalty=9999 \exhyphenpenalty=9999 }
\def\dateline{\rightline{\ifcase\month\or
  January\or February\or March\or April\or May\or June\or
  July\or August\or September\or October\or November\or December\fi
  \space\number\year}}
\def\received{\vskip 3pt plus 0.2fill
 \centerline{\sl (Received\space\ifcase\month\or
  January\or February\or March\or April\or May\or June\or
  July\or August\or September\or October\or November\or December\fi
  \qquad, \number\year)}}


\hsize=6.5truein
\hoffset=0.0truein
\vsize=8.9truein
\voffset=0.0truein
\parskip=\medskipamount
\def\\{\cr}
\twelvepoint            
\doublespace            
\overfullrule=0pt       


\newcount\timehour
\newcount\timeminute
\newcount\timehourminute
\def\daytime{\timehour=\time\divide\timehour by 60
  \timehourminute=\timehour\multiply\timehourminute by-60
  \timeminute=\time\advance\timeminute by \timehourminute
  \number\timehour:\ifnum\timeminute<10{0}\fi\number\timeminute}
\def\today{\number\day\space\ifcase\month\or Jan\or Feb\or Mar
  \or Apr\or May\or Jun\or Jul\or Aug\or Sep\or Oct\or
  Nov\or Dec\fi\space\number\year}




\def\cutp#1{
 \rightline{\rm CU--TP--#1}}


\def\title                      
  {\null\vskip 3pt plus 0.2fill
   \beginlinemode \doublespace \raggedcenter \bf}

\def\author                     
  {\vskip 3pt plus 0.2fill \beginlinemode
   \doublespace \raggedcenter}

\def\affil                      
  {\vskip 3pt plus 0.1fill \beginlinemode
   \oneandahalfspace \raggedcenter \it}

\def\abstract                   
  {\vskip 3pt plus 0.3fill \beginparmode \narrower
   \oneandahalfspace {\it  Abstract}:\  }

\def\endtopmatter               
  {\endpage                     
   \body}

\def\body                       
  {\beginparmode}               

\def\head#1{                    
  \goodbreak\vskip 0.4truein    
  {\immediate\write16{#1}
   \raggedcenter {\sc #1} \par }
   \nobreak\vskip 0truein\nobreak}

\def\beneathrel#1\under#2{\mathrel{\mathop{#2}\limits_{#1}}}

\def\refto#1{$^{#1}$}           

\def\references                 
  {\head{References}            
   \beginparmode
   \frenchspacing \parindent=0pt    
   \parskip=0pt \everypar{\hangindent=20pt\hangafter=1}}

\gdef\refis#1{\item{#1.\ }}                     

\gdef\journal#1,#2,#3,#4.{              
    {\sl #1~}{\bf #2}, #3 (#4).}                

\def\endreferences{\body}

\def\figurecaptions             
  {\endpage
   \beginparmode
   \head{Figure Captions}
}

\def\endpage                    
  {\vfill\eject}

\def\endpaper                   
  {\endmode\vfill\supereject}


\def\ref#1{Ref.~#1}                     
\def\Ref#1{Ref.~#1}                     
\def\[#1]{[\cite{#1}]}
\def\cite#1{{#1}}
\def\(#1){(\call{#1})}
\def\call#1{{#1}}
\def\taghead#1{}
\def\frac#1#2{{#1 \over #2}}
\def\half{{\frac 12}}

\def\12{{1\over2}}

\def\sla{\raise.15ex\hbox{$/$}\kern-.57em}
\def\leaderfill{\leaders\hbox to 1em{\hss.\hss}\hfill}
\def\twiddle{\lower.9ex\rlap{$\kern-.1em\scriptstyle\sim$}}
\def\bigtwiddle{\lower1.ex\rlap{$\sim$}}
\def\gtwid{\mathrel{\raise.3ex\hbox{$>$\kern-.75em\lower1ex\hbox{$\sim$}}}}
\def\ltwid{\mathrel{\raise.3ex\hbox{$<$\kern-.75em\lower1ex\hbox{$\sim$}}}}
\def\square{\kern1pt\vbox{\hrule height 1.2pt\hbox{\vrule width 1.2pt\hskip 3pt
   \vbox{\vskip 6pt}\hskip 3pt\vrule width 0.6pt}\hrule height 0.6pt}\kern1pt}
\def\tdot#1{\mathord{\mathop{#1}\limits^{\kern2pt\ldots}}}

\def\pmb#1{\setbox0=\hbox{#1}%
  \kern-.025em\copy0\kern-\wd0
  \kern  .05em\copy0\kern-\wd0
  \kern-.025em\raise.0433em\box0 }


\def\cugrant{This research was supported in part by the United States
Department of Energy under contract DE-AC02-76ER02271.}

\catcode`@=11
\newcount\tagnumber\tagnumber=0

\immediate\newwrite\eqnfile
\newif\if@qnfile\@qnfilefalse
\def\write@qn#1{}
\def\writenew@qn#1{}
\def\w@rnwrite#1{\write@qn{#1}\message{#1}}
\def\@rrwrite#1{\write@qn{#1}\errmessage{#1}}

\def\taghead#1{\gdef\t@ghead{#1}\global\tagnumber=0}
\def\t@ghead{}

\expandafter\def\csname @qnnum-3\endcsname
  {{\t@ghead\advance\tagnumber by -3\relax\number\tagnumber}}
\expandafter\def\csname @qnnum-2\endcsname
  {{\t@ghead\advance\tagnumber by -2\relax\number\tagnumber}}
\expandafter\def\csname @qnnum-1\endcsname
  {{\t@ghead\advance\tagnumber by -1\relax\number\tagnumber}}
\expandafter\def\csname @qnnum0\endcsname
  {\t@ghead\number\tagnumber}
\expandafter\def\csname @qnnum+1\endcsname
  {{\t@ghead\advance\tagnumber by 1\relax\number\tagnumber}}
\expandafter\def\csname @qnnum+2\endcsname
  {{\t@ghead\advance\tagnumber by 2\relax\number\tagnumber}}
\expandafter\def\csname @qnnum+3\endcsname
  {{\t@ghead\advance\tagnumber by 3\relax\number\tagnumber}}

\def\equationfile{%
  \@qnfiletrue\immediate\openout\eqnfile=\jobname.eqn%
  \def\write@qn##1{\if@qnfile\immediate\write\eqnfile{##1}\fi}
  \def\writenew@qn##1{\if@qnfile\immediate\write\eqnfile
    {\noexpand\tag{##1} = (\t@ghead\number\tagnumber)}\fi}
}

\def\callall#1{\xdef#1##1{#1{\noexpand\call{##1}}}}
\def\call#1{\each@rg\callr@nge{#1}}

\def\each@rg#1#2{{\let\thecsname=#1\expandafter\first@rg#2,\end,}}
\def\first@rg#1,{\thecsname{#1}\apply@rg}
\def\apply@rg#1,{\ifx\end#1\let\next=\relax%
\else,\thecsname{#1}\let\next=\apply@rg\fi\next}

\def\callr@nge#1{\calldor@nge#1-\end-}
\def\callr@ngeat#1\end-{#1}
\def\calldor@nge#1-#2-{\ifx\end#2\@qneatspace#1 %
  \else\calll@@p{#1}{#2}\callr@ngeat\fi}
\def\calll@@p#1#2{\ifnum#1>#2{\@rrwrite{Equation range #1-#2\space is bad.}
\errhelp{If you call a series of equations by the notation M-N, then M and
N must be integers, and N must be greater than or equal to M.}}\else%
 {\count0=#1\count1=#2\advance\count1
by1\relax\expandafter\@qncall\the\count0,%
  \loop\advance\count0 by1\relax%
    \ifnum\count0<\count1,\expandafter\@qncall\the\count0,%
  \repeat}\fi}

\def\@qneatspace#1#2 {\@qncall#1#2,}
\def\@qncall#1,{\ifunc@lled{#1}{\def\next{#1}\ifx\next\empty\else
  \w@rnwrite{Equation number \noexpand\(>>#1<<) has not been defined yet.}
  >>#1<<\fi}\else\csname @qnnum#1\endcsname\fi}

\let\eqnono=\eqno
\def\eqno(#1){\tag#1}
\def\tag#1$${\eqnono(\displayt@g#1 )$$}

\def\aligntag#1\endaligntag
  $${\gdef\tag##1\\{&(##1 )\cr}\eqalignno{#1\\}$$
  \gdef\tag##1$${\eqnono(\displayt@g##1 )$$}}

\def\eqalignno#1{\displ@y \tabskip\centering
  \halign to\displaywidth{\hfil$\displaystyle{##}$\tabskip\z@skip
    &$\displaystyle{{}##}$\hfil\tabskip\centering
    &\llap{$\displayt@gpar##$}\tabskip\z@skip\crcr
    #1\crcr}}

\def\displayt@gpar(#1){(\displayt@g#1 )}

\def\displayt@g#1 {\rm\ifunc@lled{#1}\global\advance\tagnumber by1
        {\def\next{#1}\ifx\next\empty\else\expandafter
        \xdef\csname @qnnum#1\endcsname{\t@ghead\number\tagnumber}\fi}%
  \writenew@qn{#1}\t@ghead\number\tagnumber\else
        {\edef\next{\t@ghead\number\tagnumber}%
        \expandafter\ifx\csname @qnnum#1\endcsname\next\else
        \w@rnwrite{Equation \noexpand\tag{#1} is a duplicate number.}\fi}%
  \csname @qnnum#1\endcsname\fi}

\def\ifunc@lled#1{\expandafter\ifx\csname @qnnum#1\endcsname\relax}

\let\@qnend=\end\gdef\end{\if@qnfile
\immediate\write16{Equation numbers written on []\jobname.EQN.}\fi\@qnend}

\catcode`@=12

\catcode`@=11
\newcount\r@fcount \r@fcount=0
\newcount\r@fcurr
\immediate\newwrite\reffile
\newif\ifr@ffile\r@ffilefalse
\def\w@rnwrite#1{\ifr@ffile\immediate\write\reffile{#1}\fi\message{#1}}

\def\writer@f#1>>{}
\def\referencefile{
  \r@ffiletrue\immediate\openout\reffile=\jobname.ref%
  \def\writer@f##1>>{\ifr@ffile\immediate\write\reffile%
    {\noexpand\refis{##1} = \csname r@fnum##1\endcsname = %
     \expandafter\expandafter\expandafter\strip@t\expandafter%
     \meaning\csname r@ftext\csname r@fnum##1\endcsname\endcsname}\fi}%
  \def\strip@t##1>>{}}

\def\citeall#1{\xdef#1##1{#1{\noexpand\cite{##1}}}}
\def\cite#1{\each@rg\citer@nge{#1}}     

\def\each@rg#1#2{{\let\thecsname=#1\expandafter\first@rg#2,\end,}}
\def\first@rg#1,{\thecsname{#1}\apply@rg}       
\def\apply@rg#1,{\ifx\end#1\let\next=\relax
\else,\thecsname{#1}\let\next=\apply@rg\fi\next}

\def\citer@nge#1{\citedor@nge#1-\end-}  
\def\citer@ngeat#1\end-{#1}
\def\citedor@nge#1-#2-{\ifx\end#2\r@featspace#1 
  \else\citel@@p{#1}{#2}\citer@ngeat\fi}        
\def\citel@@p#1#2{\ifnum#1>#2{\errmessage{Reference range #1-#2\space is bad.}
    \errhelp{If you cite a series of references by the notation M-N, then M and
    N must be integers, and N must be greater than or equal to M.}}\else%
 {\count0=#1\count1=#2\advance\count1
by1\relax\expandafter\r@fcite\the\count0,%
  \loop\advance\count0 by1\relax
    \ifnum\count0<\count1,\expandafter\r@fcite\the\count0,%
  \repeat}\fi}

\def\r@featspace#1#2 {\r@fcite#1#2,}    
\def\r@fcite#1,{\ifuncit@d{#1}          
    \expandafter\gdef\csname r@ftext\number\r@fcount\endcsname%
    {\message{Reference #1 to be supplied.}\writer@f#1>>#1 to be supplied.\par
     }\fi%
  \csname r@fnum#1\endcsname}

\def\ifuncit@d#1{\expandafter\ifx\csname r@fnum#1\endcsname\relax%
\global\advance\r@fcount by1%
\expandafter\xdef\csname r@fnum#1\endcsname{\number\r@fcount}}

\let\r@fis=\refis                       
\def\refis#1#2#3\par{\ifuncit@d{#1}
    \w@rnwrite{Reference #1=\number\r@fcount\space is not cited up to now.}\fi%
  \expandafter\gdef\csname r@ftext\csname r@fnum#1\endcsname\endcsname%
  {\writer@f#1>>#2#3\par}}

\def\r@ferr{\endreferences\errmessage{I was expecting to see
\noexpand\endreferences before now;  I have inserted it here.}}
\let\r@ferences=\references
\def\references{\r@ferences\def\endmode{\r@ferr\par\endgroup}}

\let\endr@ferences=\endreferences
\def\endreferences{\r@fcurr=0
  {\loop\ifnum\r@fcurr<\r@fcount
    \advance\r@fcurr by 1\relax\expandafter\r@fis\expandafter{\number\r@fcurr}%
    \csname r@ftext\number\r@fcurr\endcsname%
  \repeat}\gdef\r@ferr{}\endr@ferences}


\let\r@fend=\endpaper\gdef\endpaper{\ifr@ffile
\immediate\write16{Cross References written on []\jobname.REF.}\fi\r@fend}

\catcode`@=12

\citeall\refto          
\citeall\ref            %
\citeall\Ref            %

\catcode`@=11
\newwrite\tocfile\openout\tocfile=\jobname.toc
\newlinechar=`^^J
\write\tocfile{\string\input\space jnl^^J
  \string\pageno=-1\string\firstpageno=-1000\string\singlespace
  \string\null\string\vfill\string\centerline{TABLE OF CONTENTS}^^J
  \string\vskip 0.5 truein\string\rightline{\string\underbar{Page}}\smallskip}

\def\tocitem#1{
  \t@cskip{#1}\bigskip}
\def\tocitemitem#1{
  \t@cskip{\quad#1}\medskip}
\def\tocitemitemitem#1{
  \t@cskip{\qquad#1}\smallskip}
\def\tocitemall#1{
  \xdef#1##1{#1{##1}\noexpand\tocitem{##1}}}
\def\tocitemitemall#1{
  \xdef#1##1{#1{##1}\noexpand\tocitemitem{##1}}}
\def\tocitemitemitemall#1{
  \xdef#1##1{#1{##1}\noexpand\tocitemitemitem{##1}}}

\def\t@cskip#1#2{
  \write\tocfile{\string#2\string\line{^^J
  #1\string\leaderfill\space\number\folio}}}

%

\def\t@cproduce{
  \write\tocfile{\string\vfill\string\vfill\string\supereject\string\end}
  \closeout\tocfile
  \immediate\write16{Table of Contents written on []\jobname.TOC.}}


\let\t@cend=\endpaper\def\endpaper{\t@cproduce\t@cend}

\catcode`@=12

\tocitemall\head                


\def \half {\hbox {$1 \over 2$}}
\def \t {{\rm tr}}
\cutp{537}
\title{New Integrable Systems from Unitary Matrix
Models\footnote{$^*$}{\cugrant}}
\author{\bf Alexios P. Polychronakos}
\affil{Pupin Physics Laboratories, Columbia University,
New York, NY 10027}
\abstract{We show that the one dimensional unitary matrix model
with potential of the form $a U + b U^2 + h.c.$ is integrable.
By reduction to the dynamics of the eigenvalues, we establish the
integrability of a system of particles in one space dimension in an
external potential of the form $a \cos (x+\alpha ) + b \cos ( 2x +\beta )$
and interacting through two-body potentials of the inverse sine square type.
This system constitutes a generalization of the Sutherland model in the
presence of external potentials. The positive-definite matrix model,
obtained by analytic continuation, is also integrable, which leads to the
integrability of a system of particles in hyperbolic potentials interacting
through two-body potentials of the inverse hypebolic sine square type.}

\endtopmatter

\body
\baselineskip=20pt

In one space dimension an integrable class of systems
is known, involving particles coupled through two-body potentials
of a particular form. The generic type of these potentials is of the
inverse square form. Calogero [1] first solved the three-body problem
in the quantum case for inverse square interactions and quadratic external
potential. Later on, the full $N$-body problem was solved
and shown to be integrable in both the classical and the quantum case [2]
and further to be related to Lie algebras [3]. Sutherland [4] solved
the problem with inverse sine square interactions (and no external
potentials), which can be thought as the inverse square potential rendered
periodic on the circle. Eventually, it was realized that the system of
particles with two-body potentials of the Weierstrass function type is
integrable [5,6]. Again, these potentials can be thought as the inverse
square potential rendered periodic on a complex torus. For a review of
these systems and a comprehensive list of references see [6].

An interesting feature of the above systems is that, at
least some of them, admit a matrix formulation [6,7]. Specifically,
the inverse square potential arises out of a hermitian matrix
model, and the inverse sine square potential arises from a
unitary matrix model. The matrix formulation is in many respects
a better framework to study these systems. Such hermitian
matrix models have been studied in physics in the context of large-$N$
expansions [8-10] and, recently, non-perturbative two-dimensional gravity
[11]. Although the one dimensional ($c=1$) unitary model has not been
studied in this context, discrete ($c<1$) models have been considered
[12]. Further, the inverse square potential was shown to be of relevance
to fractional statistics [13] and anyon physics [14]. It becomes, therefore,
of interest to look for integrable systems with more general potentials.

In a previous paper [15] we achieved such a generalization for the
Calogero system; specifically, it was shown that the system of
particles with inverse square potential interactions remains
integrable at the presence of external potentials which are a
general quartic polynomial in the coordinate. In this paper, we
obtain a generalization of the Sutherland model; that is, we show
that the system of particles with inverse sine square interactions
remains integrable at the presence of external potentials of the
form $a \cos (x+\alpha) + b \cos (2x+\beta)$.
An appropriate scaling limit of this system, then, is shown
to reproduce the previous quartic system. In addition, the system
with all trigonometric functions in the potentials replaced
by their hyperbolic counterparts (i.e., the inverse sinh square
system with hyperbolic external potentials) is also integrable.

The fact that integrability seems to work only for the above
type of external potentials is somewhat puzzling. It is known,
for instance, through a collective field description of the hermitian
matrix model, that the large-$N$ inverse square system is integrable
for any external potential [16]. (Although there is no corresponding result
for the unitary model we have little doubt that the same is true there.)
This could be, though, just another special property of the large-$N$
system. At any rate, the method of this paper seems to work only
for the above-mentioned potentials.

The system to be considered is a unitary matrix model in one time
dimension with lagrangian
$$
{\cal L} = -\half \t ( U^{-1} \dot U )^2 - \t W(U)
\eqno(a)$$
where $U$ is a unitary $N \times N$ matrix depending on time
$t$, and overdot denotes time derivative. The potential $W(U)$
must be hermitian for consistency. The equations of motion from
\(a) read
$$
{d \over dt}( U^{-1} \dot U ) - W^{\prime} U = 0
\eqno(b)$$
Due to the invariance of \(a) under time independent unitary
transformations of $U$, there is a conserved traceless matrix, namely
$$
[ \dot U , U^{-1} ] \equiv i P
\eqno(c)$$
as can explicitly be checked using \(b). $P$ is the generator of
unitary transformations of $U$ and constitutes a kind of conserved
``angular momentum" in the (curved) space $U(N)$. The eigenvalues
of $U$, on the other hand, written as $e^{i x_n}$, $n=1, \dots N$,
can be thought as coordinates of $N$ particles on the circle.
Take, now, the particular case where all but one of the eigenvalues of
$P$ are equal, that is
$$
P = \alpha ( N |u> <u| - 1 )
\eqno(d)$$
where $|u>$ is a constant $N$-dimensional unit vector. This $P$
is naturally obtained by gauging the $U(N)$ invariance and coupling
the system to fermions [17].
Then in can be shown with a method analogous to [17] that the eigenvalues
of $U$ satisfy the equations of motion
$$
{\ddot x}_n = - V^\prime ( x_n ) + \sum_{m \neq n} {\alpha^2
\cos{x_n - x_m \over 2} \over 4 \sin^3 {x_n - x_m \over 2}}
\eqno(e)$$
where the potential of the particles is defined
$$
V(x) = W ( e^{ix} )
\eqno(f)$$
and prime denotes $x$-derivative.
These are exactly the equations of motion of particles of unit mass
moving in the
external potential $V$ and interacting through the two-body potential
$$
V_2 (x) = { \alpha^2 \over \left( 2 \sin {x \over 2} \right)^2 }
\eqno(g)$$
In the abscence of external potentials, the above system is the
Sutherland system which is known to be integrable. It is the
purpose of this paper to show that the system remains integrable
in the presence of external potentials $V(x)$ of the form
$$
V(x) = a_1 \cos x + a_2 \sin x + b_1 \cos 2x + b_2 \sin 2x
\eqno(h)$$
The above potential can be thought as a generalization on the
circle of the quartic potential on the line. Indeed, the quartic
potential is in fact a special case of the above potential and
can be obtained from it in the limit of infinite radius of the
circle. To see this, introduce explicitly the radius $R$ by
rescaling
$$
x \to {x \over R} \,\,,\,\,\,\, t \to {t \over R^2}
\eqno(i)$$
In terms of the new variables the potential becomes
$$
V(x) \to {1 \over R^2} V( {x \over R} )
\eqno(j)$$
Then, by choosing the coefficients of the potential to scale as
$$
a_1 = -{8 \over 3} R^4 a - 8 R^6 c \,\,,\,\,\,\,
a_2 = 2 R^5 b
$$
$$
b_1 = 2 R^6 c + {1 \over 6} R^4 a \,\,,\,\,\,\,
b_2 = - R^5 b
\eqno(k)$$
we see that upon taking the limit $R \to \infty$ we recover
the potential
$$
V(x) = a x^2 + b x^3 + c x^4
\eqno(l)$$
It should be clear that this is essentially the only scaling
that leads to a finite potential at the large $R$ limit. Therefore,
the integrability of the present model also contains the
integrability of the quartic system as a limiting special case.

We shall then consider the unitary matrix problem with
potential
$$
W(U) = {A \over 2} U + {A^* \over 2} U^{-1} + {B \over 2} U^2
+ {B^* \over 2} U^{-2} \,\,,\,\,\,\,
A \equiv a_1 - i a_2 \,\,,\,\, B \equiv
b_1 - i b_2
\eqno(m)$$
Through a rotation of the circle, that is, through a redefinition
of $U$ of the form $U \to e^{i\phi} U$, we can always redefine
the phase of $B$. So, for simplicity we will choose $B$ real.
We shall define the ``left-hamiltonian matrix" $H$ as
$$
H = - \half ( U^{-1} \dot U )^2 + W(U)
\eqno(n)$$
whose trace gives the energy of the system. Similarly, we
can define the ``right-hamiltonian matrix"
$$
\tilde H = - \half ( \dot U U^{-1} )^2 + W(U) = U H U^{-1}
\eqno(n1)$$
Although the trace of the above matrices is a conserved quantity,
neither of them is conserved as a matrix, nor are their eigenvalues.
Consider, now, the matrices
$$
L = H + {\kappa \over 2} [ U^{-1} \dot U , U + U^{-1} ] \,\,,\,\,\,\,
M = - \half U^{-1} \dot U + {\kappa \over 2} ( U - U^{-1} )
\eqno(o)$$
Then, upon using the equations of motion it is possible to show that
$$
\dot L + [ L , M ] = 0
\eqno(p)$$
provided that we choose $\kappa$ as
$$
\kappa^2 = B
\eqno(q)$$
Therefore, the matrices $L$ and $M$ constitute a Lax pair [18] and
the eigenvalues of $L$ are conserved. Equivalently, the traces
$$
I_n = \t L^n
\eqno(r)$$
are constants of the motion. Restoring, now, an arbitrary phase
to $B$, through a phase redefinition of $U$, we get that the
Lax matrices in the most general case are
$$
2 L= - \half ( U^{-1} \dot U )^2 + A U + B U^2 + \kappa [ U^{-1} \dot U ,
U ] + h.c.
\eqno(s)$$
$$
M = - \half U^{-1} \dot U + \kappa U - h.c.
\eqno(s1)$$
where $h.c.$ denotes hermitian conjugate. Notice that $L$ is
hermitian while $M$ is antihermitian. Relations \(p) and \(q)
remain unchanged.

Alternatively, we could write the ``right-Lax pair" matrices
$\tilde L$, $\tilde M$, by substituting $\dot U U^{-1}$ for
$U^{-1} \dot U$ in \(s) and \(s1) and flipping the sign of the first
term in $M$. The two choices are really equivalent, connected through
a unitary transformation generated by $U$ itself.

The above are true without any assumptions for the constant commutator $P$.
Assuming, now, that $P$ is of the form \(d), we can express the
conserved quantities $I_n$ in terms of particle coordinates $x_n$ and
momenta $p_n = {\dot x}_n$. To see this, perform a (time-dependent)
unitary transformation on $U$ which brings it to a diagonal form
while it doesn't change $I_n$. Then, using \(d), we see
that the matrices of the system take the form
$$
U_{ij} = \delta_{ij} e^{i x_j} \,\,,\,\,\,\, P_{ij} =
\alpha ( \delta_{ij} - 1 )
$$
$$
( U^{-1} \dot U )_{ij} = i {\dot x}_i \delta_{ij} +
{ i \alpha ( 1 - \delta_{ij} ) \over e^{i( x_i - x_j )} - 1}
\eqno(u)$$
The $I_n$, therefore, can be expressed using \(u) in terms of
particle quantities and constitute $N$ independent integrals of motion
for the system of particles in the potential \(h) interacting through
inverse sine square two-body potentials of strength $\alpha^2$. $I_1$
is the total energy of the system. $I_2$ has the form
$$
I_2 = \sum_i \left( \half p_i^2 + \sum_{j \neq i} {\alpha^2 \over
2 s_{ij}^2} + V( x_i ) \right)^2 + \sum_{i \neq j \neq k \neq i} \Bigl(
{\alpha^2 \over 2 s_{ij} \, s_{jk}} \Bigr)^2
$$
$$
+ \sum_{i \neq j} \left( {\alpha ( p_i + p_j ) \over 2 s_{ij}}
\right)^2 - {|B| \alpha^2 \over 2} \sum_{i \neq j} \cos ( x_i + x_j +
\beta ) \eqno(u1)$$
where we defined
$$
s_{ij} = 2 \sin { x_i - x_j \over 2} \,\,,\,\,\,\,
B = |B| e^{i \beta}
\eqno(u2)$$
and used the identity
$$
\sum_{\{i,j,k,l\,\,{\rm distinct}\}} {1 \over s_{ij} \, s_{jk} \, s_{kl}
\, s_{li} } = 0
\eqno(u3)$$
We shall not give here the explicit form
of the rest of $I_n$ in terms of $x_i$, $p_i$, which is quite
complicated. Their independence is obvious from the fact that, just as
in the hermitian case, $I_n$ is a polynomial in $p_i$ of order $2n$
with highest order term of the form $\sum_i p_i^{2n}$, and such terms
cannot be obtained from lower order terms for $n \leq N$.

The above Lax pair matrices go over to the Lax matrices of the
hermitian problem in the limit of infinite radius, under the
simultaneous scaling \(i) and \(k) for the coordinates and the
potential, and the rescaling
$$
L \to {1 \over R^2} L \,\,,\,\,\,\, M \to {1 \over R^2} M
\eqno(t)$$
In the case $c=0$, though, there is no scaling limit, expressing
the fact that the purely cubic potential does not admit a Lax pair
formulation although it is still integrable.

It should also be noted that, just as in the hermitian case, there are
two distinct Lax pairs, corresponding to the two possible choices of sign
for $\kappa$ in \(q). Again, this has no impact on the conserved
quantities $I_n$ since the two choices give essentially the same set
of conserved quantities, modulo traces of $P^n$, which are trivial
nondynamical constants in the space of eigenvalues. Note also that,
in the hermitian case, when the coefficient of the quartic term
became negative the Lax matrix $L$ became non-hermitian, indicating
the runaway nature of the system. The integrals of motion, nevertheless,
remained real. In the present case no such thing happens, since
there can be no runaway behavior on the circle, and the Lax matrix
\(s) remains hermitian for all values of the coefficients of the potential.

The special case $B=0$ is interesting: it corresponds to the quadratic
hermitian model (i.e., the Calogero system) which is obtained as the
large $R$ scaling limit in this case. In particular, $L=H$ just as in
the hermitian case. Contrary to the Calogero system, however, where $H$
is a constant matrix, in this case there is still some nontrivial
unitary rotation with time, due to the nonzero curvature of the space.
Therefore, the gauge matrix of the Lax pair
$M = - \half U^{-1} \dot U$, generating this rotation, does not
vanish. At the scaling limit, of course, $M$ goes to zero.

It remains to show the fact that the above quantities are in involution,
that is, that their Poisson brackets vanish. Again, as in the
hermitian case, trying to use the explicit expressions of $I_n$
in terms of particle phase space variables and using
$$
\{ x_i , p_j \} = \delta_{ij}
\eqno(v)$$
is completely hopeless. Instead, we shall work with the canonical
structure of the original matrix problem and make use of the fact
that the projection from the full matrix phase space of $U$ to
the phase space of its eigenvalues is a hamiltonian reduction [6,7].
To see this, define from \(a) a canonical momentum $P_U$
$$
P_U = {\delta {\cal L} \over \delta {\dot U}} = - U^{-1} \dot U U^{-1}
\eqno(w)$$
Then the symplectic two-form $\omega$ is
$$
\omega = \t \, ( d P_U \, dU )
\eqno(x)$$
giving rise to canonical Poisson brackets for $U$ and $P_U$. For
convenience, we shall work with the antihermitian matrix
$$
\Pi \equiv - P_U U = U^{-1} \dot U
\eqno(y)$$
and the symplectic one-form $\cal A$, in terms of which we have
$$
{\cal A} = - \t \, ( \Pi \, U^{-1} dU ) \,\,,\,\,\,\, \omega = d {\cal A}
\eqno(z)$$
Decompose, now, $U$ and $\Pi$ in terms of diagonal and angular
degrees of freedom, namely
$$
U = V^{-1} \Lambda V \,\,,\,\,\,\,
\Pi = V^{-1} \Bigl( N + \Lambda^{-1} [ A , \Lambda ] \Bigr) V
\eqno(aa)$$
with $\Lambda$ and $N$ diagonal unitary and antihermitian matrices,
respectively:
$$
\Lambda_{ij} = e^{x_i} \delta_{ij} \,\,,\,\,\,\,
N_{ij} = i p_i \, \delta_{ij}
\eqno(bb)$$
and $A$ an off-diagonal antihermitian matrix which will be
identified with the off-diagonal part of $\dot V V^{-1}$ by the
equations of motion. Then, in this parametrization \(z) becomes
$$
\eqalign{
{\cal A} &= \t \,\Bigl( - N \Lambda^{-1} d \Lambda + ( \Lambda A
\Lambda^{-1} + \Lambda^{-1} A \Lambda - 2A ) \, dV V^{-1} \Bigr) \cr
&= \sum_i p_i \, dx_i + \t \Bigl( i P V^{-1} dV \Bigr) \cr}
\eqno(cc)$$
So $\cal A$ decomposes into two non-mixing parts, the first one
being the particle phase space one-form leading to \(v) and
the second part identifying $P$ as the variable canonically
conjugate to the angular degrees of freedom. Therefore, the
Poisson brackets of quantities expressible in terms of $x_i$
and $p_i$ can equally well be evaluated using the full matrix
phase space Poisson structure.

To show the involution of $I_n$ it is much more convenient to use
the geometric formulation of symplectic manifolds, rather than work
explicitly with Poisson brackets. In this approach, we map to each
function $f$ on the phase space a vector field $v_f$ on the phase
space through the relation
$$
< v_f , \omega > = df
\eqno(dd)$$
where $< \,,\, >$ denotes internal product (contraction).
The matrix differential operators $\delta_\Pi$ and $\delta_U U$ form a
basis for the vector fields, satisfying
$$
< ( \delta_\Pi )_{ij} , d\Pi_{kl} > = < ( \delta_U U )_{ij} ,
( U^{-1} dU )_{kl} > = \delta_{il} \delta_{jk} \,\,,\,\,\,\,
{\rm else}\,\,\, < \,.\,,\,.\, > = 0
\eqno(ee)$$
In this basis, we can express the vector field $v_f$  as
$$
v_f = \t ( v^1 \delta_\Pi + v^2 \delta_U U )
\eqno(ff)$$
Then the Poisson bracket of any two functions $f$, $g$ can be
calculated as
$$
\{ f , g\} = v_f (g) = \t \Bigl( v^1 {\delta g \over \delta \Pi}
+ v^2 {\delta g \over \delta U} U \Bigr)
\eqno(gg)$$
Therefore, defining $v_n = v_{I_n}$, we have
$$
\{ I_n , I_m \} = v_n (I_m) = m \, \t \, \Bigl( L^{m-1} v_n (L) \Bigr)
\eqno(ii)$$
On the other hand, $v_n$ satisfies
$$
< v_n , \omega > = d I_n = \t ( L^{n-1} dL )
\eqno(jj)$$
By expressing $L$ in \(s) in terms of $\Pi$ and $U$ and expanding
$dL$ as
$$
dL = \sum_i R_i \, d\Pi \, S_i + \sum_i {\tilde R}_i \, U^{-1} dU
\, {\tilde S}_i
\eqno(kk)$$
and using the expression for $\omega$ derived from \(z), that is
$$
\omega = \t \Bigl( - d\Pi U^{-1} dU + \Pi ( U^{-1} dU )^2 \Bigr)
\eqno(ll)$$
we get from \(jj)
$$
v_n = -n \sum_i \t \Bigl( {\tilde S}_i L^{n-1} {\tilde R}_i \delta_\Pi
+ [ S_i L^{n-1} R_i , \Pi ] \delta_\Pi - S_i L^{n-1} R_i \delta_U U
\Bigr)
\eqno(mm)$$
Finally, substituting \(mm) in \(ii) we obtain
$$
\{ I_n , I_m \} = -nm \sum_{i,j} \t \Bigl( L^{m-1} R_i {\tilde S}_j
L^{n-1} {\tilde R}_j S_i + L^{m-1} R_i S_j L^{n-1} R_j \Pi S_i \Bigr)
- ( m \rightleftharpoons n )
\eqno(nn)$$
The rest is a tedious algebraic exercise, calculating the
explicit expressions for $R_i$, $S_i$, ${\tilde R}_j$ and
${\tilde S}_j$ and evaluating \(nn). We shall omit all detail
and simply give the result
$$
\{ I_m , I_n \} = mn ( \kappa^2 - B ) \, \t \, \Bigl( L^{m-1} U
L^{n-1} [ \Pi , U ] \Bigr) \,+\, h.c. \,\,- ( m \rightleftharpoons n)
\eqno(oo)$$
Choosing then $\kappa$ as in \(q) the Poisson brackets \(oo) vanish
and the $I_n$ are in involution. We have therefore proved the full
integrability of the system.

The expression for $L$ can be brought to a particularly suggestive
form in a somewhat special case. Define the matrices
$$
A = \Pi + Y(U) \,\,,\,\,\,\, A^{\dagger} = - \Pi + Y(U) \,\,,\,\,\,\,
{\rm with} \,\, Y(U) = \kappa U + \kappa^* U^{-1} + \lambda
\,\,,\,\,\,\, \lambda = {\rm real}
\eqno(pp)$$
Then $L = \half A^\dagger A$ is exactly the previous Lax matrix \(s),
with $B$ and $\kappa$ properly connected. The coefficients of the
potential, however, are constrained to obey the relation
$$
{B \over A^2} = {\rm real}
\eqno(qq)$$
The situation is again analogous to the hermitian case, where
such an expression for $L$ also existed for a similarly constrained
potential. Note that, in this case
$$
I_n (\kappa , \lambda ) = \t ( \half A^\dagger A )^n =
\t ( \half A A^{\dagger} ) = I_n (-\kappa , -\lambda )
\eqno(rr)$$
which shows that the two possible choices of sign
for $\kappa$ in \(q) give exactly the same conserved quantities.
The proof of involution of the $I_n$ is also simplified in
this case.

The above results can readily be extended to the case of positive
definite matrices $X$. Such matrices can be expressed as
$$
X = e^M \,\,,\,\,\,\, M \,\,{\rm hermitian.}
\eqno(ss)$$
They differ therefore from unitary matrices by a mere analytic
continuation, and can be thought as unitary models with imaginary
radius $R$. We can explicitly perform this continuation in the
expressions for the Lax pair matrices, by putting $R=i$ in \(i),
\(j),\(k) and \(t), which amounts to the redefinitions
$$
U \to X \,\,,\,\,\,\, L \to - L
$$
$$a_1 \pm i a_2 \to -( a_1 \pm a_2 )
\,\,,\,\,\,\, b_1 \pm i b_2 \to - ( b_1 \pm b_2 )
\eqno(tt)$$
and now $\kappa$ and $\kappa^*$ are distinct, satisfying
$$
\kappa^2 = -b_1 - b_2 \,\,,\,\,\,\, {\kappa^*}^2 = -b_1 + b_2
\eqno(uu)$$
The new potential of the system becomes
$$
V(x) = a_1 \cosh x + a_2 \sinh x + b_1 \cosh 2x + b_2 \sinh 2x
\eqno(vv)$$
Notice that, for \(vv) to be bounded from below, we must
have
$$
b_1 \pm b_2 > 0
\eqno(ww)$$
and for such values $\kappa$ and $\kappa^*$ both become imaginary.
The Lax matrix becomes, explicitly
$$
L = \half \Pi^2 + W(X) + i \kappa_1 [ \Pi , X ]
+ i \kappa_2 [ \Pi , X^{-1} ]
\eqno(ww1)$$
where we put $\kappa = 2i \kappa_1$, $\kappa^* = 2i \kappa_2$.
The momentum $\Pi = X^{-1} \dot X$ now is not antihermitian any
more, but rather satisfies
$$
\Pi^\dagger = X \Pi X^{-1}
\eqno(xx)$$
Therefore, although $L$ is not hermitian, it satisfies
$$
L^\dagger = X L X^{-1}
\eqno(yy)$$
which ensures that the conserved quantities $I_n$ are real.
Alternatively, we can directly make the analytic continuation
$$
x_i \to i x_i \,\,,\,\,\,\, p_i \to i p_i
\eqno(zz)$$
into the expressions for $I_n$ and obtain the new expressions.
These will be conserved quantities for particles on the line,
in an external potential of the type \(vv) interacting through
two-body potentials of the type
$$
V_2 (x) = {\alpha^2 \over \left( 2 \sinh {x \over 2} \right)^2}
\eqno(aaa)$$
Systems of particles with the above two-body potential and no external
potentials were known to be integrable.  As we proved, they remain
integrable in the presence of potentials of the form \(vv).

In conclusion, we see that the above unitary system (and its positive
definite analytic continuation) is integrable and closely parallels the
hermitian case. It is interesting that integrability again
stops at potentials with only their two lowest nontrivial Fourier
coefficients nonvanishing and does not extend to arbitrary periodic
potentials. Obvious attempts to generalize the result to higher
potentials, e.g., through a construction similar to \(pp) with
arbitrary hermitian $Y(U)$, fail. This may be just a shortcoming of the
method used. If, on the other hand, it turns out that integrability
indeed stops at this level, it would be interesting to understand
the deeper reason for that and the special significance of these
potentials.

We should stress that, just as in the hermitian case,
integrability does not necessarily imply complete solvability. The
significance of the results is that the problem in the presence of
the interparticle interaction remains as solvable as in the case
of decoupled particles. The solution of these systems could
conceivably be achieved in terms of the solutions of the one-body
problem (which in general involves elliptic functions).
The proof of integrability in the quantum domain is also of interest,
especially in view of the conjecture put forth in [13] that the
inverse square two-body potential on the line and the inverse
sine square potential on the circle simply endow the particles
with fractional statistics. These, as well as possible generalizations
for other two-body potentials (e.g., of the Weierstrass form)
remain interesting topics for further work.

\bigskip
\bigskip
\centerline {\bf REFERENCES}

\noindent
\item{[1]}
F. Calogero, {\it J. Math. Phys.} {\bf 12} (1971) 419.

\item{[2]}
J. Moser, {\it Adv. Math.} {\bf 16} (1975) 1; F. Calogero, {\it Lett.
Nuovo Cim.} {\bf 13} (1975) 411; F. Calogero and C. Marchioro, {\it Lett.
Nuovo Cim.} {\bf 13} (1975) 383.

\item{[3]}
M.A. Olshanetsky and A.M. Perelomov, {\it Invent. Math.} {\bf 37}
(1976) 93.

\item{[4]}
B. Sutherland, {\it Phys. Rev.} {\bf A4} (1971) 2019 and {\bf A5}
(1972) 1372.

\item{[5]}
B. Sutherland, {\it Phys. Rev. Lett.} {\bf 34} (1975) 1083.

\item{[6]}
M.A. Olshanetsky and A.M. Perelomov, {\it Phys. Rep.} {\bf 71} (1981) 314
and {\bf 94} (1983) 6.

\item{[7]}
D. Kadzan, B. Kostant and S. Sternberg, {\it Comm. Pure Appl.
Math.} {\bf 31} (1978) 481.

\item{[8]}
A. Jevicki and H. Levine, {\it Phys. Rev. Lett.} {\bf 44} (1980) 1443;

\item{[9]}
G. Marchesini and E. Onofri, {\it J. Math. Phys.} {\bf 21} (1980)
1103.

\item{[10]}
M.B. Halpern, {\it Nucl. Phys.} {\bf B188} (1981) 61; M.B. Halpern
and C. Schwartz, {\it Phys. Rev.} {\bf D24} (1981) 2146;
M.B. Halpern, {\it Nucl. Phys.} {\bf B204} (1982) 93.

\item{[11]}
D.J. Gross and I.R. Klebanov, {\it Nucl. Phys.} {\bf B344} (1990)
475 and {\bf B354} (1991) 459.

\item{[12]}
Representative works in a growing literature on the subject are:
V. Periwal and D. Shevitz, {\it Phys. Rev. Lett.} {\bf 64}
(1990) 1326 and {\it Nucl. Phys.} {\bf B344} (1990) 344; K. Demeterfi
and C.-I Tan, {\it Mod. Phys. Lett.} {\bf A5} (1990) 1563; C. Crnkovic,
M. Douglas and G. Moore, {\it Nucl. Phys.} {\bf B360} (1991) 507;
M.J. Bowick, A. Morozov and D. Shevitz, {\it Nucl. Phys.} {\bf B354}
(1991) 496.

\item{[13]}
A.P. Polychronakos, {\it Nucl. Phys.} {\bf B324} (1989) 597.

\item{[14]}
J.M. Leinaas and J. Myrheim, {\it Phys. Rev.} {\bf B37} (1988) 9286;
A.P. Polychronakos, {\it Phys. Lett.} {\bf B264} (1991) 362.

\item{[15]}
A.P. Polychronakos, Columbia preprint CU-TP-527, July 1991.

\item{[16]}
J. Avan and A. Jevicki, SLAC/Brown preprint SLAC-PUB-BROWN-HET-801.

\item{[17]}
A.P. Polychronakos, Columbia preprint CU-TP-520, to appear in
Phys. Lett. B.

\item{[18]}
P. Lax, {\it Comm. Pure Appl. Math.} {\bf 21} (1968) 467.

\end